\documentclass[12pt,letterpaper]{JHEP3}
\usepackage{cite}
\usepackage{epsfig}

\title{A paradox in the global description of the multiverse}

\author{%
Raphael Bousso and Ben Freivogel\\ 
Center for Theoretical Physics, Department of Physics\\
University of California, Berkeley, CA 94720-7300, U.S.A.\\
{\em and}\\
Lawrence Berkeley National Laboratory, 
Berkeley, CA 94720-8162, U.S.A.}
\abstract{%
We use an argument by Page to exhibit a paradox in the global
description of the multiverse: the overwhelming majority of observers
arise from quantum fluctuations and not by conventional evolution.
Unless we are extremely atypical, this contradicts observation.  The
paradox does not arise in the local description of the multiverse,
but similar arguments yield interesting constraints on the maximum
lifetime of metastable vacua.
}

\preprint{\hepth{0610132}}

\begin{document}

\section{Introduction}

The landscape of string theory may explain the cosmological constant
problem, but it has also given rise to new challenges.  Of these
perhaps the most formidable is our lack of techniques for making
predictions in a theory with $10^{500}$ vacua.  This task may require
a drastic revision of our picture of the universe on the largest
scales.

The standard, ``global'' picture is that of an eternally inflating
``multiverse'', containing an infinite number of bubbles, or pocket
universes, corresponding to each vacuum in the landscape.  Each bubble
is an infinite open universe, and if it has positive cosmological
constant, it will itself harbor an infinite number of daughter
universes.
\EPSFIGURE{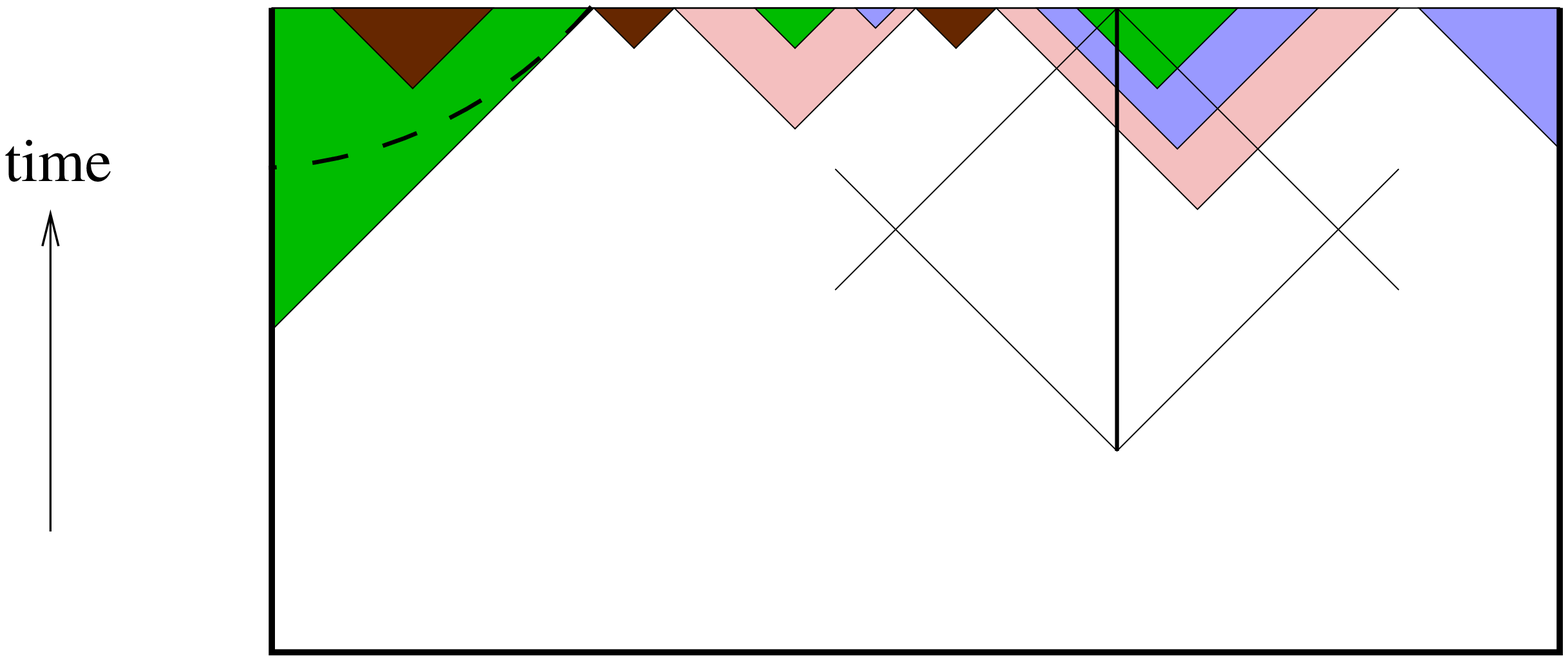,width=.8\textwidth}{\label{fig-global}
  Globally, the universe contains regions corresponding to every
  vacuum in the landscape (shown in different colors).  Each such
  region is an infinite open universe; the dashed line shows an
  example of an open equal-time hypersurface.  The black diamond is an
  example of a spacetime region that is causally accessible to a
  single observer.  Note that the global universe is not visible to
  any one observer.}

The new, ``local'' picture is minimalist: It consists only of the
spacetime region causally accessible from one worldline.  This region
has the shape of a causal diamond~\cite{Bou00a}, defined as the
overlap between the past light-cone of a late-time event with the
future light-cone of an early event on the worldline.  This is what
one observer can probe, and {\em this is all there is\/}, as far as the
semiclassical description of the universe goes.

As long as a theory can deal with {\em any\/} worldline and the
associated causal diamond, it can describe all experiments that can be
done in a semiclassical geometry.  Thus, it is clearly sufficient to
restrict to this region, and the economy of it may appeal to some.
But is it necessary?  What is wrong with the more intuitive, global
point of view?  What possible harm could it do?

In this paper, we first review known difficulties of the global point
of view in section \ref{sec-probs}. While these difficulties are
worrisome, they are not obviously fatal for the global description of
cosmology.  In section \ref{sec-pp} we use an argument recently given
by Page to exhibit a novel paradox: In the global description of the
multiverse, almost all observers arise from random fluctuations,
rather than by conventional evolution.  This would make us extremely
atypical and thus conflict with observation.  In section \ref{sec-res}
we discuss how the local viewpoint resolves the Page paradox.
However, similar considerations yield a much weaker but nontrivial
constraint on the lifetime of all vacua in the landscape that can
contain observers.

The paradox discussed here should be regarded as analogous to the
quantum xerox paradox in black hole physics. Just as in black hole
physics, we believe the resolution lies in abandoning the global
description.

\section{Problems of the global picture}
\label{sec-probs}

\subsection{A predictivity crisis}
\label{sec-amb}

We would like to predict low energy physics parameters observers are
likely to observe.  This requires statistical sampling of the theory
landscape; an understanding of how the cosmological dynamics favors or
disfavors the production of each vacuum; and finally, a sensible
method for estimating the abundance of observers in each vacuum.  (For
example, parameters unique to a vacuum with no observers have zero
probability of being observed.)

But in the global picture, it is not clear how to regulate the
infinite number of infinitely large bubbles, so as to compare the
number (or volume?) of regions corresponding to different
vacua~\cite{LinLin94}.  A number of proposals have been made,
e.g.~\cite{GarVil01,GarSch05,EasLim05}.  But we lack a convincing
principle that would tell us which, if any, is correct.
(Diffeomorphism invariance is sometimes cited, but any function of a
such a measure will share the same property.)

In our view, this crisis of predictivity does not arise from a lack of
sophistication or ingenuity in the prescriptions that have been
proposed.  It is the global point of view itself that is at fault.  We
are trying to regulate infinities that are figments of our
imagination, and struggling to reign in volumes that cannot be seen by
anyone without violating causality.

\subsection{The FHW problem}
\label{sec-fhw}

In the global picture, it is natural to include volume expansion
factors in the probability of vacua.  All other things being equal, a
vacuum that harbors a long period of slow-roll inflation (say, a
billion $e$-foldings) will be $e^{10^9}$ times more likely than one
with only $60$ $e$-foldings.

Feldstein, Hall, and Watari~\cite{FelHal05} have pointed out that this
leads to a problem, because in generic families of models of
inflation, the duration of inflation either grows or shrinks
monotonically with the size of the density perturbations produced.
Thus $\delta\rho/\rho$ would be driven to $0$ or $1$ with exponential
preference.  Anthropic arguments~\cite{TegRee97} cannot help
here---most observers would result from the exponential tails of
Gaussian distributions and would find themselves alone in a hostile
universe.  But this is not what we see; the universe is full of
galaxies like ours.  Indeed, $\delta\rho/\rho\approx 10^{-5}$ is
comfortably inside the anthropic window.

To be fair, this problem could be an artifact of an overly naive
sampling of inflationary models.  But the exponentially large volume
factors make it difficult to come up with a credible sample of models
such that $\delta\rho/\rho$ will be small but not exponentially small.

\subsection{The quantum-xeroxing paradox for black holes}

There is now considerable evidence that black hole formation and
evaporation can be described as a unitary process by an outside
observer.  Yet, this would appear to create a paradox. 

There is an instant of time at which both the collapsing star inside
the black hole, and the Hawking radiation that allegedly carries away
its quantum state, are in regions of negligible curvature, where
semiclassical gravity should be valid.  But this would mean that the
quantum state of the star has been copied: it exists both inside and
outside the black hole.  This violates the linearity of quantum
mechanics.

The paradox is resolved~\cite{SusTho93,Pre92} by noting that no
observer can actually verify this violation.  Either the observer
stays outside, seeing only the Hawking radiation; or the observer
falls in and sees only the star.  There is no observer whose causal
past can include both copies of the quantum state.

This suggests that the global viewpoint must be abandoned to avoid
severe inconsistencies.  It is not clear how such a conclusion can be
confined to the context of black holes.  Rather, one would expect it
to apply generally, and thus in particular to cosmology.  

However, no comparable, sharp paradox has been shown to plague the
global description in the context of cosmology.  We will now use a
recent observation by Page to fill this gap.

\section{The Page Paradox}
\label{sec-pp}

In a long-lived vacuum with positive cosmological constant, structure
can form in two ways. Structure can form in the conventional way (through a
period of inflation followed by reheating), or it can form spontaneously
as a rare thermal fluctuation. Because de Sitter space is thermal, if
the vacuum is sufficiently long-lived spontaneous structure formation 
will occur.

Observation indicates that the structure we see today was formed by
conventional means. As explained by Dyson, Kleban, and Susskind
\cite{DysKle02}, if structure forms spontaneously it is exponentially
unlikely to be describable by a sensible semiclassical history. For
example, it is exponentially more likely for a single galaxy to
fluctuate into existence than for the observed universe to form as a
fluctuation. Among observers who form from thermal fluctuations, the
vast majority will be close to the smallest fluctuation which can
constitute an observer. This is the ``Boltzmann's Brain''
paradox---within this framework, most observers are isolated brains
which fluctuate from the vacuum in the absence of any other structure.

Page avoids the difficult problem of comparing different vacua in the
multiverse and asks the following question: in the vacuum we find
ourselves in, do more observers form conventionally or spontaneously?
Since we seem to be in a spatially infinite universe, there are an
infinite number of both types of observers. Page focuses on a finite
comoving volume to regulate the spatial infinity. In a given comoving
volume, a finite number of conventional observers form.\footnote{Some
particularly rare fluctuations will happen to reproduce conventional
evolution from a hot big bang, but they are exponentially less
frequent than the production of isolated observers in an empty
universe~\cite{DysKle02}.} However, if our vacuum decays slowly enough
that it eternally inflates, then the undecayed physical volume
continues to grow with time. As a result, an infinite number of
observers form spontaneously in a finite comoving volume. Page
concludes that the decay rate of our vacuum must be fast enough that
inflation is not eternal; otherwise, we would be infinitely atypical
observers.  The required decay time to avoid eternal inflation is of
order the time scale set by the cosmological constant,
$\Lambda^{-1/2}$, i.e., of order $10^{10}$ years.

It is not absurd to suggest that our vacuum will decay in a few
billion years. However, as Page points out, his analysis suggests a
stronger conclusion. If \it any \rm vacuum which is capable of
supporting observers eternally inflates, such a vacuum produces an
infinite number of Boltzmann brains in a finite comoving
volume. Presumably, this infinity would imply that a typical observer
in the multiverse is a Boltzmann brain, and we would have to conclude
that no vacuum capable of supporting observers eternally
inflates. Since this stronger conclusion depends on comparing the
relative probability of different vacua, other infinities could
conceivably arise which would avoid this conclusion. We suspect,
however, that any formulation of probabilities which relies on a
global point of view will lead to the following conclusion: The
observation that we observe conventional structure formation, together
with the assumption that we are typical, implies that no vacuum
capable of harboring observers eternally inflates.

Such a conclusion would be shocking, and is at odds with our current,
admittedly crude, understanding of the string landscape.  For example,
in the toy model of Ref.~\cite{BP}, our vacuum was estimated to have a
lifetime of order $\exp(10^{10})$ if the number of fluxes is $O(100)$.
With more fluxes, the upper exponent can be somewhat decreased but it
would need to decrease to $3$ for the lifetime to become of order ten
billion years.  This cannot be accomplished with a realistic number of
fluxes if we still wish to solve the cosmological constant problem---a
key motivation to consider the landscape in the first place.

In particular, we can show that the proposal of \cite{GarSch05}, which
we consider the state of the art in globally inspired probability
measures, suffers from the Page Paradox.  In this proposal, the
probability of measurements is computed in two steps. First, each
vacuum is assigned an \it a priori \rm probability $P_i$. The $P_i$
encode the dynamics of eternal inflation, and they are finite. In the
second step, each vacuum is weighted by the number of observers within
a unit comoving volume. The probability of observing vacuum $i$ is
proportional to the product of the a priori probability and the number
of observers per comoving volume.

Page's argument shows that if inflation is eternal, and the
cosmological constant is small enough to fit one observer within the
cosmological horizon, then the number of observers per comoving
volume is infinite. So as long as at least one eternally inflating,
observer-allowing  vacuum exists with a finite a priori probability,
the final probability distribution is zero for any vacuum which does
not eternally inflate. Furthermore, in the eternally inflating
vacua, observers are infinitely more likely to be Boltzmann brains
than honest folk like ourselves.

One is left with a clear choice: either eternally inflating vacua
admitting observers are shockingly absent from the string landscape,
or the proposal \cite{GarSch05} is incorrect. Though we cannot prove
it, we expect the same difficulty to arise in any globally inspired
proposal which has finite a priori probabilities for vacua with positive
cosmological constant.

There is a deeper underlying reason for this problem.  The global
picture is, in a sense, an expansion about the least likely worldlines
(those which fail to enter terminal vacua for an atypically long
time).  From the global viewpoint, the extreme unlikeliness of a
worldline's evolution is more than compensated by the exponentially
large volume expansion factor it picks up.  Hence, the global geometry
of eternal inflation is dominated by regions which arose from the most
unlikely evolution.  Then it should not surprise us that the majority
of observers can be similarly characterized, and arise from highly
unlikely fluctuations.

\section{A resolution: The local viewpoint}
\label{sec-res}

In the local viewpoint, the universe consists only of one (any)
causally connected region of causal-diamond form.  In our vacuum, for
example, this region is the interior of the de~Sitter horizon.
(Strictly, it is overlap of the above ``top cone'' with the interior
of a future lightcone which can be taken to start at reheating, but
this restriction imposed by the bottom cone will not be needed here.)

In a vacuum with positive cosmological constant, the de~Sitter horizon
is finite, with area of order $\Lambda^{-1}$.  Thus, the number of
observers will be finite at any time.  We can still be concerned about
Boltzmann brains, and it remains true that we expect the first
Boltzmann observers to show up after an exponentially long but finite
time of order 
\begin{equation}
t_{BB}\sim \exp(ER)~,
\end{equation}
 where $E$ is the energy of the brain and
$R=\Lambda^{-1/2}$ is the radius, and thus the inverse temperature, of
the de~Sitter space.  (This time is the inverse of the Boltzmann
factor, up to negligible prefactors.  It can also be obtained directly
from the entropy decrease of the heatbath, the cosmological horizon,
when an object of energy $E$ forms in de~Sitter space~\cite{Bou00b}.)

The difference is that in the global picture, your last chance to
destroy the vacuum and prevent Boltzmann brains was much earlier, at a
time of order $R$~\cite{Page}.  This is because the universe is
exponentially expanding after that time, and a bubble of new vacuum
cannot catch up and completely replace the old vacuum.  

In the local picture, the causal diamond is all there is.  No-one can
go and probe the exponentially large regions allegedly created by the
cosmological expansion, so we do not consider them to be part of
reality.  What remains---the causal diamond---has constant asymptotic
size $R$\footnote{Vacua with negative cosmological constant also have
finite causal diamonds because they contain a big crunch. In the case
of false vacua which can decay to $\Lambda=0$, one might think an
infinity appears even in the local analysis. Since such vacua must be
supersymmetric, they cannot contain observers.}  It can be easily
replaced in its entirety by a new vacuum.  Once a bubble forms, it
will quickly expand out to the cosmological horizon.

Hence, all that we need is for the vacuum to decay before the first
Boltzmann brains start appearing, after a time $t_{\rm B}$.\footnote{This
  conclusion, as well as the discussion below, arose in e-mail
  discussions with T.~Banks in July 2006.}  That is, we need its
lifetime to be shorter than $\exp(ER)$.  This is an exponentially
weaker constraint than the lifetime of order $R$ required in the
global picture.

Nevertheless, the need to purge Boltzmann brains imposes interesting
constraints on the landscape of string theory.  It would seem to be
important to avoid Boltzmann brains in all vacua.  This includes
de~Sitter spaces with larger temperature, and thus higher brain
creation rate.  Because $T \sim 1/R$, the highest temperature
is set by the smallest possible size of the Boltzmann brain: $R>R_{\rm
  B}$.  For example, with $E=100$ kg and $R=1$ m Boltzmann would be
suppressed by $\exp(-ER)\sim \exp(-10^{45})$ (up to a negligible
entropy factor).  This relatively high rate applies not in our
universe, but in some other vacuum with high enough cosmological
constant to allow Boltzmann to fit snugly inside the cosmological
horizon.

It seems plausible\footnote{The basic idea is that the problem arises
  only in vacua with cosmological constant small compared to 1, or
  else not enough entropy will fit to admit an observer, Boltzmann or
  otherwise~\cite{Bou00a}.  But to achieve a small cosmological
  constant requires combining a number of ingredients that can lead to
  accidental cancellation~\cite{BP}.  Hence there will be a large
  number of decay channels.  It would be surprising if none of them
  satisfies Eq.~(\ref{eq-bb}), especially since each ingredient should
  come from a fundamental theory and thus be tied to high energy
  scales.}, but not obvious, that all vacua in the string landscape
would decay sufficiently fast.  The relevant comparison is the
instanton action for the {\em fastest\/} decay channel of each vacuum,
vs.~the $ER$ of a Boltzmann brain that could form in it.  The decay
would have to win this competition in all vacua.  In other words, we
need that the time to nucleate a Boltzmann brain is longer than the
lifetime of the false vacuum,
\begin{equation}
t_{BB} > t_{decay} ~,
\label{eq-bb}
\end{equation}
for all vacua in the landscape.

It is unclear how to characterize the requirements for an object to be
a Boltzmann brain. One way of looking at it is that an ordered object
must fluctuate out of the thermal bath of de Sitter space.  That is,
de Sitter space must fluctuate into a lower entropy configuration. The
difference in entropy is a measure of how many computations the
``observer" can perform before melting back into the heat
bath~\cite{Bou06}.  For now, we do not try to quantify the necessary
entropy difference, $S_{BB}$, but for an intelligent observer it must
be quite large.

The expected amount of time to create such an entropy difference scales as
\begin{equation}
t_{BB} \sim e^{S_{BB}} \ .
\end{equation}
This time is very large. On the other hand, the typical decay time for
a metastable vacuum is also exponentially large, since the decay is
nonperturbative. As a result, (\ref{eq-bb}) is a nontrivial constraint
on the decay time, but a much weaker constraint than Page's.

For similar reasons (large inflationary expansion factors do not
change the asymptotic size of the causal diamond), the local viewpoint
also resolves the FHW problem (Sec.~\ref{sec-fhw}).  This is discussed
in Ref.~\cite{Bou06}, where is is also argued that unambiguous
probabilities are obtained in this approach (Sec.~\ref{sec-amb}).

We would like to thank A.~Aguirre, T.~Banks, M.~Lippert, M.~Kleban,
L.~Susskind and I.~Yang for discussions.  We are especially indebted
to T.~Banks who first impressed upon us that the Boltzmann paradox is
a potential concern even for a landscape with terminal vacua.

\bibliographystyle{board}
\bibliography{all}
\end{document}